\documentclass[11pt,twoside]{article}
\usepackage{asp2010}

\resetcounters

\begin{document}

\title{March of the Starbugs: Configuring Fibre-bearing Robots on the UK-Schmidt Optical Plane}
\author{Nuria~P.~F.~Lorente$^1$, Minh~Vuong$^1$, Christophe~Satorre$^{1,2}$, Sungwook~E.~Hong$^3$,
Keith~Shortridge$^1$, Michael~Goodwin$^1$ and Kyler~Kuehn$^1$
\affil{$^1$Australian Astronomical Observatory, PO Box 915, North Ryde, NSW 1670, Australia}
\affil{$^2$Laboratoire de Syst\`emes Robotiques (LSRO), Ecole Polytechnique F\'ed\'erale de Lausanne
  (EPFL), Station 9, 1015 Lausanne, Switzerland}
\affil{$^3$Department of Physics, Korea Institute for Advanced Study, Seoul 130-722, Korea}}

\begin{abstract}
The TAIPAN instrument, currently being developed for the Australian Astronomical Observatory's UK
Schmidt telescope at Siding Spring Observatory, makes use of the AAO's Starbug technology to deploy
150 science fibres to target positions on the optical plane. This paper describes the software
system for controlling and deploying the fibre-bearing Starbug robots. The TAIPAN software is
responsible for allocating each Starbug to its next target position based on its current position
and the distribution of targets, finding a collision-free path for each Starbug, and then
simultaneously controlling the Starbug hardware in a closed loop, with a metrology camera used to
determine the position of each Starbug in the field during reconfiguration. The software is written
in C++ and Java and employs a DRAMA middleware layer~\citep{1995fbs}.
\end{abstract}

\section{Introduction}
Multi Object fibre Spectroscopy (MOS) is a well-established technique for efficiently carrying out
spectroscopy on a large number of targets in the field of view. The AAO's 2dF instrument
\citep{2002lct+} has facilitated spectroscopic studies of many thousands of objects in the 17 years
since it began operation --- e.g. 2dFGRS \citep{2001cdm+}, 2QZ \citep{2000bsc+}, WiggleZ
\citep{2010djb+}, GAMA \citep{2011dhk+} and GALAH \citep{2015dfb+}.

Due to its single robot arm and the sequential nature of its operation, the field configuration time
of 2dF increases linearly with the number of fibres. 
This and other limitations (such as diversity of payload and non-planar focal planes) 
are resolved by the AAO's new Starbugs technology, which consists of one independently positionable
robot per science fibre. The initial phase of TAIPAN, the first instrument to make use of this
technology, will have 150 science fibres and therefore consist of 150 Starbugs. This will allow the
field configuration process to be carried out in parallel, and thereby decrease the configuration
time from around 60~min for 2dF to the order of 5 minutes.

\section{The TAIPAN Instrument and Survey}
A Starbugs positioner is currently being developed for the TAIPAN instrument \citep{2014klb+} on the AAO's 1.2m,
6° FoV UK-Schmidt telescope, located at Siding Spring Observatory in New South Wales, Australia, and
is scheduled to commence on-sky observations in early 2016.

The TAIPAN survey will obtain visible band spectra for $5\times10^5$ Southern Sky galaxies (${\sim}70\%$
completeness) at R=2200 in $14<r<18$ at SNR=5--10 (30~min exposures).  The related FunnelWeb stellar
survey will obtain spectra for $2\times10^6$ Southern stars (${\sim}99\%$ completeness) in $5.7<v<12$ at
SNR=100 (30 min exposures)

The primary scientific aims of the 5-year survey include providing a measurement of $H_0$ to within 2\%
and improving by a factor of 2 the measured accuracy of the local growth rate, resulting in
stronger tests of General Relativity.  Additionally TAIPAN will carry out a precision peculiar
velocity survey and engage in studies of galaxy evolution, transition, environment and fuelling studies.

TAIPAN will also serve as a prototype instrument for MANIFEST (Many Instrument Fibre System)
\citep{2014lbb+} on the Giant Magellan Telescope (GMT), scheduled for completion in 2021.

\section{Starbugs}
A Starbug \citep{2014bcg+} consists of two concentric piezo-electric ceramic tubes which are made to move
over a glass field plate by the alternate deformation of the inner and outer tubes. This is done by
applying a voltage in a given sequence to generate a "walking" motion, with
which the Starbug is positioned to within a few microns of the target. The 8-mm diameter Starbug
carries an optical fibre payload and three back-illuminated metrology fibres. Adhesion to the field
plate is provided by a vacuum system which keeps the Starbug on the plate whilst allowing freedom of
motion.

\section{The TAIPAN Positioner Software System}

The positioner software is responsible for open-- and closed--loop control and monitoring of the
150~Starbugs on the field plate. It determines the current location of each Starbug, assigns its
next target position, calculates a valid trajectory and sends the appropriate control commands via
the electronics firmware layer to move it to its new position. Secondary to this
the software system collects and displays monitor information to allow the user to ascertain the
health of the instrument. Finally, the software also stores monitoring data for the long-term
characterisation and evaluation of the instrument's performance. The software system comprises the
following modules:

\begin{itemize}
\item Master Controller: responsible for starting up and shutting down the software system and
  ensuring that all components are operational. It also looks after the archiving of Starbug
  operational properties and monitor data.

\item Positioner: this is responsible for the motion of the 150 Starbugs, both in a field
  reconfiguration (expected to occur on average once per hour and take ${\sim}5$ min) and in position
  adjustment during an observation. Reconfiguration is carried out in closed-loop control, with the
  Metrology module providing position updates on the Starbugs whilst they are in motion.
  Given a target field (a list of sky positions) the Positioner allocates a Starbug to each
  target position, determines a route for each Starbug and iteratively moves the Starbugs to their
  new positions while receiving position updates from the metrology module, until the Starbugs are
  within the target tolerance.

\item Metrology: this controls the 29M-pixel camera which images the Starbugs' back-illuminated metrology
fibres and processes the resulting $6576 \times 4384$ pixel images to find the position of the science
fibre for each Starbug, based on the measured position of its 3 metrology fibres.
Together with the Positioner, the Metrology module forms part of the Starbug control closed
loop, which keeps track of Starbug locations at a given time, detects lost Starbugs (e.g. if a bug's
metrology fibres fail to illuminate, a bug falls off the plate due to vacuum failure, etc.)
or unusable Starbugs
(e.g.\ one which does not respond to movement commands, has faulty metrology fibres or has
otherwise been marked as bad by the system). This module is also crucial in Starbug initialisation
and position calibration.

\item User Interface:
will provide two degrees of control over the system, for engineering and observing uses, and
give operational feedback.

\item Instrument Simulator:
simulates the behaviour of the Starbugs, the firmware layer
and the Metrology system. It allows the software to be tested with no hardware
present.
\end{itemize}

\section{Route Finding and Position Allocation}
For each Starbug a collision-free path must be found which minimises the time to target, the amount
of Starbug rotation (to avoid focal ratio degradation in the science fibre) and the tangling of
fibres between Starbugs. Additionally the path assigned to a Starbug must not render another's target
position unreachable.

Because optimum path determination can be computationally expensive for some field configurations,
route finding for the TAIPAN instrument is designed to be carried out either ``live'' during the
plate reconfiguration process, or off-line (e.g.\ during the day) allowing one to set up a series of
observations ahead of time. The off-line process calculates 3 separate routes for each Starbug: 1)
the path between the current (configuration n-1) and next (configuration n) targets; 2) the path
between the current target and the home position; and 3) the path between the home position and the
next target. Because there is always a pre-calculated path to the Starbugs' park position
recovering from a fault or skipping one or more configurations (e.g.\ due to bad weather) can be
done quickly and with a minimum of path recalculation.

Although the Starbugs can be controlled independently, due to constraints imposed by the electronics
system the movement mode (rotation or translation) must be the same for all Starbugs driven by a
given electronics rack. Because of this, the routing software divides the reconfiguration process
into a series of Ticks, each defined as an interval (not necessarily of equal duration) during which
all the Starbugs either rotate on their central axes, translate, or wait.

The positioner uses three stages of increasing complexity in determining a valid path for each
Starbug. If a path cannot be found using the earlier (computationally cheaper) methods, an attempt
is made with the more complex and expensive methods until a valid path is found or the target is
flagged as unreachable:

\begin{enumerate}
\item {\bf Simple Vector:} This is a priority-based positioner. It detects crossings and possible
  collisions between pairs of Starbug paths in the current Tick, and prioritises those with no
  crossings, calculating their paths first. Once these simplest paths are done the positioner
  calculates the path for one of the Starbugs of each crossing pair, then the second of each pair
  and finally any remaining routable bugs. Unroutable bugs progress to stage 2.

\item {\bf Traffic Light:} This introduces the option for a Starbug to wait at a position along its
  path for one or more Ticks, so as to avoid a collision. As before, unroutable bugs progress to
  stage 3.

\item {\bf Traffic Light + Cooperative $A^*$:} The $A^*$ algorithm determines the minimum-cost path
  using a combination of the geometric (past) cost and the heuristic (future) cost. However, it is
  grid-based (we require continuous positioning of Starbugs on the field plate) and does not avoid
  moving objects (the Starbugs themselves are dynamic obstacles).  We are using a modification of
  this algorithm, the Cooperative $A^*$, which adds a time dimension to avoid the paths of other
  Starbugs by reserving a path through the grid in a given Tick. The grid used is 4-dimensional
  consisting of 2 cartesian axes, a Tick axis and a reserved path axis. This approach finds a path
  for a significant number of Starbugs, but is expensive, and so is only used for Starbugs which are
  otherwise unroutable. 
\end{enumerate}

\noindent Although a successful strategy in a large number of use-cases, tests so far show that some
cases remain unresolved by this 3-phased approach. Work on further refining the Traffic
Light + Cooperative $A^*$ method continues, together with improvements to the algorithm which
allocates Starbugs to target positions, with the aim of avoiding the occurrence of deadlocks and
improve overall efficiency.

\bibliography{P4-4}

\end{document}